\documentclass[twocolumn]{aastex61}

\usepackage{hyperref}
\usepackage{graphicx}
\usepackage{bigstrut}

\newcommand{\ledd}{$L_{\rm Edd}$}
\newcommand{\src}{GX 5$-$1}
\newcommand{\nustar}{{\it NuSTAR}}
\newcommand{\swift}{{\it Swift}}
\newcommand{\nh}{$N_{\rm H}$}
\newcommand{\fe}{Fe K$_\alpha$}

\shorttitle{No reflection in GX 5$-$1}
\shortauthors{Homan et al.}

\begin{document}

\title{Absence of reflection features in \nustar\ spectra of the luminous neutron star X-ray binary \src}

\correspondingauthor{Jeroen Homan}
\email{jeroen@space.mit.edu}

\author{Jeroen Homan}
\affiliation{Eureka Scientific, Inc., 2452 Delmer Street, Oakland, CA 94602, USA}
\affiliation{SRON, Netherlands Institute for Space Research, Sorbonnelaan 2, 3584 CA Utrecht, The Netherlands}
\affiliation{MIT Kavli Institute for Astrophysics and Space Research, 70 Vassar Street, Cambridge, MA 02139, USA}

\author{James F.\ Steiner}
\affiliation{MIT Kavli Institute for Astrophysics and Space Research, 70 Vassar Street, Cambridge, MA 02139, USA}
\affiliation{Einstein Fellow}
\author{Dacheng Lin}
\affiliation{Space Science Center, University of New Hampshire, Durham, NH 03824, USA}
\author{Joel K.\ Fridriksson}
\affiliation{Anton Pannekoek Institute, University of Amsterdam, Science Park 904, 1098 XH, Amsterdam, The Netherlands}
\author{Ronanld A.\ Remillard}
\affiliation{MIT Kavli Institute for Astrophysics and Space Research, 70 Vassar Street, Cambridge, MA 02139, USA}
\author{Jon M.\ Miller}
\affiliation{Department of Astronomy, University of Michigan, 1085 South University Ave, Ann Arbor, MI 48109-1107, USA}
\author{Renee M.\ Ludlam}
\affiliation{Department of Astronomy, University of Michigan, 1085 South University Ave, Ann Arbor, MI 48109-1107, USA}





\begin{abstract}

We present \nustar\ observations of the luminous neutron star low-mass X-ray binary (NS LMXB) and Z source \src. During our three observations made with separations of roughly two days, the source traced out an almost complete Z track. We extract spectra from the various branches and fit them with a continuum model that has been successfully applied to other Z sources. Surprisingly, and unlike most of the (luminous) NS-LMXBs observed with \nustar, we do not find  evidence for reflection features in any of the spectra of \src. We discuss several possible explanations for the absence of reflection features. Based on a comparison with other accreting neutron star systems and given the high luminosity of \src\ ($\sim$1.6--2.3 times the Eddington luminosity, for a distance of 9 kpc), we consider a highly ionized disk the most likely explanation for the absence of reflection features in \src.

\end{abstract}


\section{Introduction} \label{sec:intro}

The most luminous neutron star low-mass X-ray binaries (NS-LMXBs), the so-called Z sources, trace out characteristic multi-branched tracks in their X-ray color--color and hardness--intensity diagrams (HIDs). The origin of these branches remains poorly understood, but given the high luminosity of the Z sources --- near or above the Eddington luminosity (\ledd)  ---  radiation pressure is believed to play an important role \citep{chhaba2006,lireho2009}. Based on the shape of their tracks, two subclasses of the Z sources are generally recognized: the Sco-like and the Cyg-like Z sources \citep{kuvaoo1994}. From transient and highly variable sources,  we know that Cyg-like Z-source behavior is associated with higher luminosities \citep{hovafr2010,frhore2015}.

Various models have been proposed to fit the continuum spectra of the Z sources \citep[e.g.][]{chhaba2006,lireho2009,tisesh2014}. While most  provide statistically acceptable fits,  they often result in different interpretations of Z-source behavior (e.g.\ evolution of the inner disk radius, boundary layer area, and mass accretion rate). Reflection features in the X-ray spectra \citep[e.g.,][]{rofayo1999} could shed additional light on the changes in the accretion flow along the Z track. They are the result of irradiation of the inner accretion disk and are sensitive to changes in the accretion geometry. In black hole (BH) X-ray binaries and low-luminosity NS-LMXBs the irradiating flux is supplied by a hot corona \citep[e.g.,][]{funoto2015,sltoki2016,wamesa2017}, while in the high-luminosity NS-LMXBs, the boundary layer is the main contributor \citep[e.g.,][]{mipafu2013,lumiba2017}. 

The reflection process results in two main  spectral features \citep{rofayo1999}:  an \fe\ line around  6.4--7.0 keV (due to fluorescence) and (less frequently) a broad hump between 10--30 keV (due to Compton backscattering). Modeling of these features can provide valuable information on the geometry of the accretion flow \citep{faro2010} and changes therein.  The relativistically broadened \fe\ line, in particular, is frequently used to measure inner disk radii, in order to determine black hole spins \citep[see][for a review]{re2014} or constrain the neutron star equation of state \citep[e.g.,][]{camibh2008,camiba2010}. In view of the uncertainties in modeling the Z-source spectra, the \fe\ line is also valuable in that it could provide  independent constraints on the disk and boundary layer.

\nustar\ observations of 13 NS-LMXBs have been published, including two Sco-like Z sources \citep[GX 17+2 and GX 349+2;][]{lumiba2017,cocami2017} and one Cyg-like one \citep[Cyg X-2;][]{modepa2018}. As can be seen from Table \ref{tab:overview}, \fe\ lines were found in 11 of these sources. Here, we present an analysis of three \nustar\ observations of \src, a Cyg-like Z source \citep{kuvaoo1994}. The aim of these observations was to use inner disk measurements obtained from the reflection component as a check for the inner disk behavior obtained from our continuum modeling (see \citealt{cocami2017} and \citealt{modepa2018}, for similar attempts). However, no reflection features were detected in the spectra of \src. In Section \ref{sec:obs}, we present our observations and data analysis; in Section \ref{sec:results}, the results of our analysis;  and in Section \ref{sec:disc} we discuss possible explanations for the absence of reflection features in \src.

\begin{deluxetable*}{cccl}
\tablecaption{Neutron star LMXBs observed with \nustar \label{tab:overview}}
\tablehead{\colhead{Source} & \colhead{$L$/\ledd\tablenotemark{a}} & \colhead{\fe\ line?} & \colhead{References} }
\startdata
Cen X-4                & 3.8$\times10^{-6}$ & no & \citet{chtogr2014} \\ 
4U 1608$-$52     & 0.01--0.02 & yes &  \citet{demich2015} \\ 
4U 1636$-$536    & 0.01, 0.03--0.06 & yes  & \citet{lumiba2017},  \citet{wamesa2017}\\ 
1RXS J180408.9$-$34205 & 0.03--0.04, 0.1 & yes   &  \citet{dealpa2016}, \citet{lumica2016} \\
XTE J1709$-$267   & 0.04--0.06 & yes &  \citet{lumica2017} \\
4U 1728$-$34      & 0.08, 0.06--0.09 & yes  & \citet{sltoki2016}, \citet{mopade2017} \\ 
GS 1826$-$238    & 0.13 & no  & \citet{chgain2016} \\ 
4U 1705$-$440   & 0.2 & yes  & \citet{lumiba2017} \\ 
Ser X-1                & 0.6 & yes  & \citet{mipafu2013}, \citet{madiia2017}   \\ 
Aql X-1                 & 0.14--0.7 & yes & \citet{kitomi2016}, \citet{lumide2017} \\ 
GX 349+2          & 0.4--0.7\tablenotemark{b} & yes & \citet{cocami2017} \\
GX 17+2             & 1.2 & yes  &  \citet{lumiba2017} \\ 
Sco X-1               & 1.3 & yes &  This work\\ 
Cyg X-2              & 1.8\tablenotemark{c} & yes & This work, \citet{modepa2018} \\ 
GX 5$-$1           & 1.6--2.3 & no & This work \\ 
\enddata
\tablenotetext{a}{\ledd\ defined as 1.8$\times10^{38}$ erg\,\,s$^{-1}$.}
\tablenotetext{b}{Assuming 5 kpc \citep{chsw1997}.}
\tablenotetext{c}{Assuming 11 kpc \citep{gamuha2008}.}
\end{deluxetable*}


\section{Observations and data analysis}\label{sec:obs}

\subsection{\nustar}

\src\ was observed three times with \nustar: on 2016 April 9, 11, and 13 (ObsIDs 30102056002, 30102056004, and 30102056006). The observations had dead-time-corrected exposures of 10.3--11.2 ks. The reason for spacing the observations $\sim$2 days apart was to increase the chances of covering as much of the Z track as possible with relatively short observations. 
The  data were analyzed with the NuSTARDAS subpackage of HEASOFT V6.19, using calibration files from 2016 December. Each observation was  reprocessed with {\tt nupipeline}. Source spectra and light curves were extracted from the FPMA/FPMB detectors  using a circular region with a 120\arcsec\ radius, centered on the source. For background spectra and light curves, a similar extraction region was used, located $\sim$8\arcmin--9\arcmin\ away from the source. 

To construct a HID, we used {\tt nuproducts} to extract source/background light curves with time bins of 100 s in two energy bands:  5.6--9.6 keV and 9.6--20 keV. Live-time, point-spread-function, exposure, and vignetting corrections were applied. Background-corrected light curves from FMPA and FMPB were added, and the resulting light curves for the two bands were summed to produce a broadband light curve that is used as the intensity in the HID. The ratio of the second and first energy bands was used as the hardness. 

Because the source showed strong evolution in its HID, spectra were extracted based on location along the Z track to minimize the effects of mixing different spectral shapes. Parts of the HID were selected manually and appropriate good-time intervals were produced for each selection. Spectra were extracted with {\tt nuproducts} using these good-time intervals. The FPMA/FPMB spectra were fitted simultaneously, and a constant (with fitted values between 0.97 and 0.98) was introduced in our models to account for cross-calibration differences. 
Spectral channels below 3 keV and above 50 keV were excluded and the remaining channels were grouped to a minimum of 30 counts per spectral bin. Spectral fits were made with XSPEC V12.9.1 \citep{ar1996}. The errors quoted in this paper are all 1$\sigma$.

\swift\ observations made quasi-simultaneously with our first and second observations suffered from strong pile-up and a soft excess below 1 keV that is commonly seen in heavily absorbed sources. Due to problems in cross-calibration with the \nustar\ spectra, the \swift\ spectra were not included in our analysis.



\section{Results}\label{sec:results}

\subsection{Hardness-Intensity Diagram}

\begin{figure}[t]
\centerline{\includegraphics[width=8.5cm]{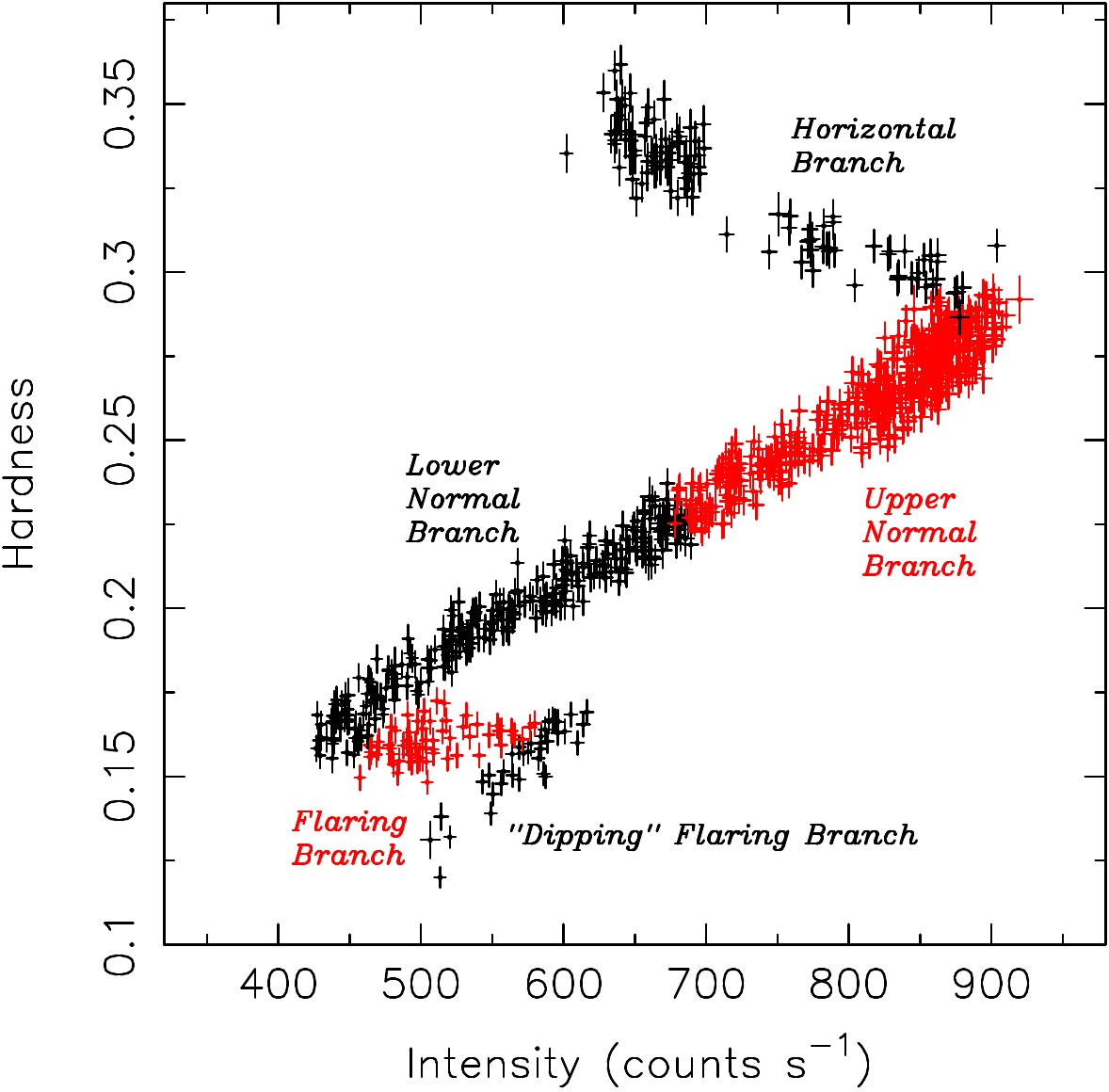}}
\caption{Combined \nustar\ HID of \src. The various Z-source branches are labeled and colored black/red to distinguish selections used for spectral analyses. Data point represent 100 s time intervals.}\label{fig:hid}
\end{figure}

In Figure \ref{fig:hid} we show the combined HID of our three \nustar\ observations of \src. An almost complete Z track was traced out. We have labeled and colored the various Z-source branches. During the first observation, the source traced out the flaring branch (FB), the ``dipping" FB, and the lower part of the normal branch (NB). The full NB was traced out during the second observation, while the upper NB and the  horizontal branch (HB) were traced out during the third observation. Gaps in the track, most notable on the ``dipping" FB and the HB, are the result of Earth occultations and passages through the South Atlantic Anomaly.

\subsection{Spectra}

FPMA/FPMB spectra were extracted from five locations along the Z track in the HID, corresponding to the black and red colored regions in Figure \ref{fig:hid}.  We note that  spectra for the lower NB were only extracted from observation 1 and upper NB spectra only from observation 2, even though these parts of the Z track were also partially traced out in other observations.

\begin{figure}[t]
\centerline{\includegraphics[width=8.5cm]{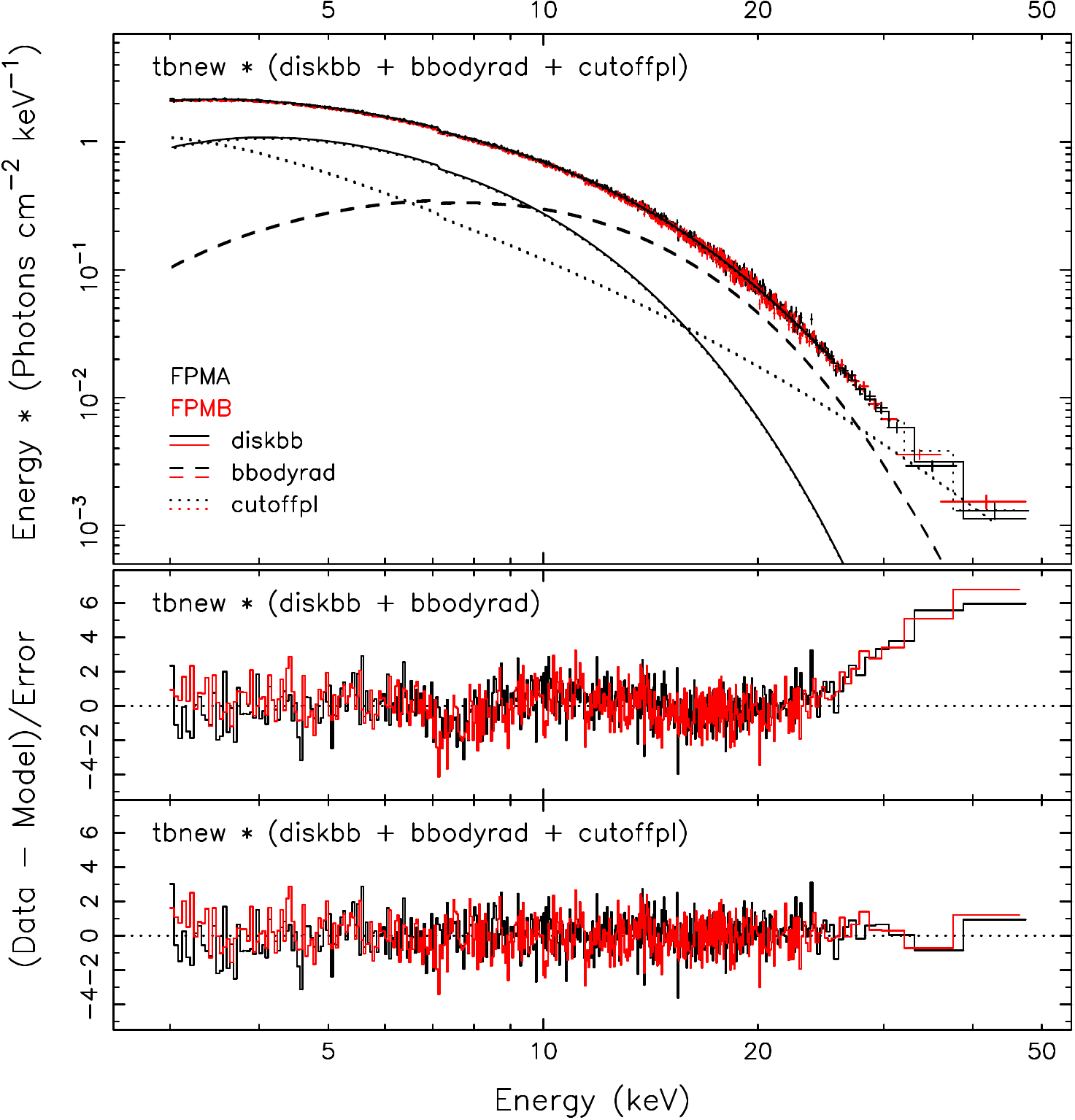}}
\caption{Fit to the HB spectra with our final model (top), and residuals from fits with our final model (bottom) and a model without a cutoff power law (middle). } \label{fig:spectrum}
\end{figure}

Following \citet{lireho2009}, we started by fitting the spectra with a continuum model consisting of a multi-color disk blackbody ({\tt diskbb} in XSPEC) and a blackbody ({\tt bbodyrad}). We used the {\tt tbnew\_feo} model  \citep{wialmc2000} for interstellar absorption, with the abundances set to {\tt wilm} and cross sections set to {\tt vern} \citep{vefeko1996}.  The values of \nh\ were left free for each selection. Although this model generally provided good fits to the continua below $\sim$15--25 keV, strong positive residuals were seen above that energy, while negative residuals around 7 keV were seen in some spectra as well (see Fig.\ \ref{fig:spectrum}, middle panel).  This model resulted in $\chi^2_{\rm red}/d.o.f.$ = 1.39/4698.  The addition of a power-law component ({\tt pegpwrlw}), as  clearly detected in {\it INTEGRAL} spectra of \src\ \citep{paebti2005}, improved the high-energy residuals, as well as those around 7 keV, and resulted in $\chi^2_{\rm red}/d.o.f.$ = 1.15/4693. However,  the power-law indices were very high, increasing from 3.6 on the HB to 4.5 on the FB, before sharply dropping to 2.1 on the dipping FB. Such steep power laws can result in degeneracies between \nh, the {\tt diskbb} parameters, and the power-law index. To lessen these degeneracies, we replaced the power law with a cutoff power law ({\tt cutoffpl}). We additionally tied the power-law index between all observations, but left the cutoff energy to vary. We also tied the \nh\ values between the spectra, as it was otherwise found to increase substantially on the FB and dipping FB. Although these constraints did not significantly improve the fit ($\chi^2_{\rm red}/d.o.f.$ = 1.14/4697), the resulting power-law index was considerably lower: 3.02$\pm$0.12. Physically motivated Comptonization models, such as, e.g., {\tt nthcomp} \citep{zdjoma1996,zydosm1999}, did not provide  better fits than our model with the cutoff power law, nor did they solve the issues with the high power-law indices. An example fit with our final model to the HB spectra is shown in Figure \ref{fig:spectrum}.

 In Table \ref{tab:fits} we report the fit results with our final model. Some general trends can be seen: as the source moves  from the HB to the dipping FB the temperatures of the disk and blackbody ($kT_{\rm disk}$ and $kT_{\rm bb}$, respectively) decrease monotonically, while their radii increase (quite dramatically so on the FB and dipping FB). The power-law contribution to the 3--50 keV flux shows an overall decrease along the Z track, while the cutoff shows a moderate decrease in energy ($E_{\rm cut}$); it was poorly constrained on the dipping FB and fixed to the FB value. The broadband unabsorbed luminosity ($L_{\rm 1-100}$), obtained by extrapolating\footnote{The power-law contribution was only measured in the 3--100 keV range} the model fits down to 1 keV and up to 100 keV and assuming a distance of 9 kpc ($d_{9}$=1), varies between $\sim3.0\times10^{38}$ erg\,\,s$^{-1}$ and  $\sim4.3\times10^{38}$ erg\,\,s$^{-1}$, peaking toward the upper NB. For a 1.4 $M_\odot$ neutron star, these luminosities correspond to Eddington ratios of $\sim$1.6--2.3($d_{9}$)$^2$. 

\begin{deluxetable*}{cccccc}[t]
\tablecaption{Spectral-fit parameters for model {\tt tbnew\_feo $\times$ (diskbb + bbodyrad + cutoffpl)}  \label{tab:fits}}
\tablehead{ \colhead{} & \colhead{HB} & \colhead{Upper NB} & \colhead{Lower NB} & \colhead{FB} & \colhead{Dipping FB}} 
\startdata
$kT_{\rm disk}$ (keV)                                        		& 1.99$\pm$0.06 & 1.83$\pm$0.02 & 1.45$\pm$0.03 & 0.98$\pm$0.02     & 0.80$\pm$0.01 \\   
$R_{\rm in}$ (km)                                               		& 8.1$\pm$0.5   & 12.4$\pm$0.4  & 17.8$\pm$0.7  & 38$\pm$3          & 70$\pm$3 \\  
$kT_{\rm bb}$ (keV)                                          		& 2.51$\pm$0.04 & 2.46$\pm$0.03 & 1.89$\pm$0.03 & 1.480$\pm$0.007   & 1.425$\pm$0.003 \\
$R_{\rm bb}$ (km)                                           		& 4.1$\pm$0.3   & 3.7$\pm$0.2   & 6.5$\pm$0.4   & 16.7$\pm$0.3      & 22.8$\pm$0.2 \\
$E_{\rm cut}$ (keV)                                         		& 18$\pm$3      & 13.1$\pm$1.6  & 10.9$\pm$1.2   & 8.2$\pm$1.0       & 8.2 (fixed) \\     
$F_{\rm 3-50\,keV}$$^{\rm a}$  ($10^{-8}$ erg\,cm$^{-2}$\,s$^{-1}$)	                & 2.12$\pm$0.02 & 2.46$\pm$0.01 & 1.80$\pm$0.01 & 1.70$\pm$0.01     & 1.85$\pm$0.01 \\      
PL Fraction (\%)    	                                                 & 29$\pm$5  & 23$\pm$5  & 26$\pm$6   & 15$\pm$5       & 1.1$\pm$1.4 \\
$L_{\rm 1-100\, keV}$ ($10^{38}$ erg\,s$^{-1}$)       & 2.98$\pm$0.01 & 3.85$\pm$0.02 & 3.15$\pm$0.02 & 3.44$\pm$0.02      & 4.25$\pm$0.02 \\ 
\enddata 

\tablecomments{The best-fit \nh\ value and cutoff power-law index were (6.20$\pm$0.16)$\times10^{22}$ atoms\,cm$^{-2}$ and 3.02$\pm$0.12, respectively. For $R_{\rm in}$, $R_{\rm bb}$, and $L_{\rm 1-100}$ we assumed a distance of 9 kpc \citep{chsw1997}. For $R_{\rm in}$  we additionally assumed an inclination of 60$^\circ$. For $L_{\rm 1-100\, keV}$ the power-law contribution was only measured between 3 and 100 keV.}
$^{\rm a}$ Unabsorbed
\end{deluxetable*}

\begin{figure}[t]
\centerline{\includegraphics[width=6.5cm]{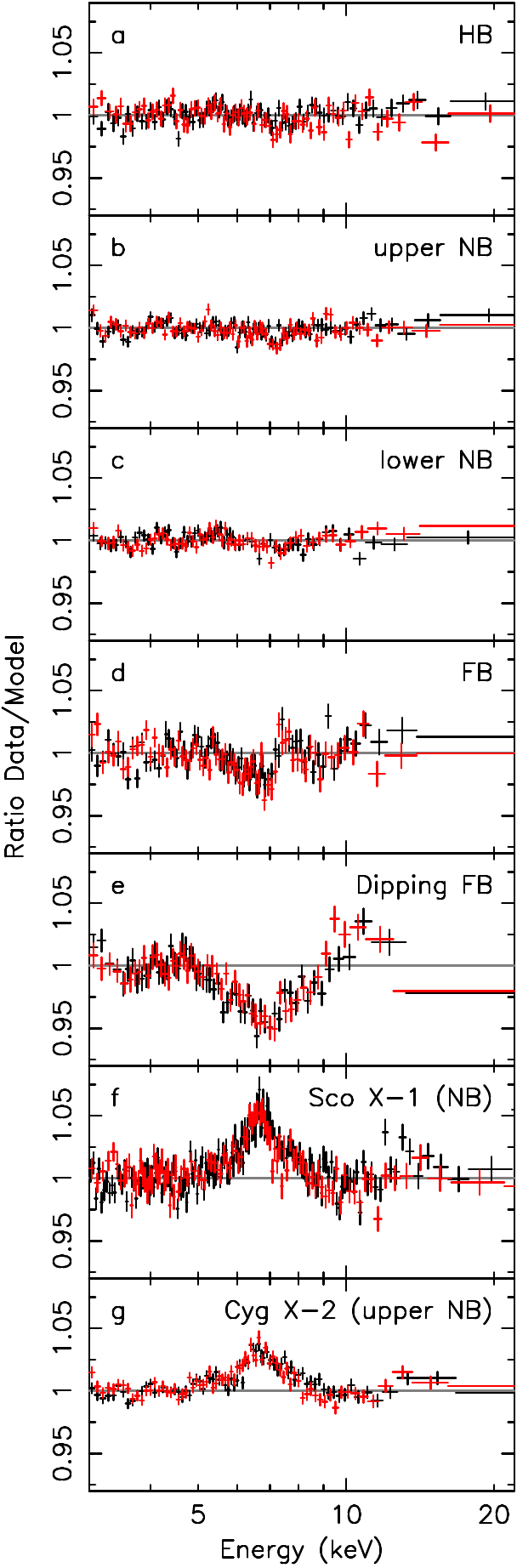}}
\caption{Data-to-model ratios for the spectra from our five HID selections and for archival \nustar\ spectra of Sco X-1 (panel (f)) and Cyg X-2 (panel (g)). Data in the 5.0--8.0 keV were initially excluded from the fit, but were reintroduced for the plot. FMPA data are shown in black, FPMB data in red. }\label{fig:ratios}
\end{figure}

The lack of coverage below 3 keV may lead to considerable (systematic) uncertainties in the continuum parameters, in particular the disk temperatures/radii, and the \nh\ values.  One of the main goals of our observations was to use reflection features, as seen in other luminous NS-LMXBs, to obtain independent measures of the evolution of the inner disk radius along the Z track. However, we do not see any indications for an \fe\ line  or a Compton hump in our spectra of \src. To illustrate this, we followed \citet{lumiba2017} and excluded data in  the 5.0--8.0 keV range in fits to the spectra of each individual HID selection. All parameters were left free and the cutoff power-law index and \nh\ were no longer tied between selections, to limit possible bias. In Figure \ref{fig:ratios} (panels (a)--(e)) we plot data-to-model ratios resulting from these fits, but with data from the 5.0--8.0 keV range included again to look for indications of (broad) \fe\ lines.  None of the spectra show positive residuals, which would indicate the presence of such lines. We do, however, see strong negative residuals in the FB and dipping FB spectra. Although negative residuals could indicate the presence of absorbing material in our line of sight (e.g.\ caused by a wind outflow), the width of the features (i.e., broader than typically seen for wind lines) might alternatively suggest that they result from shortcomings in our continuum model on those branches.

\begin{deluxetable*}{ccc}[t]
\tablecaption{Spectral-fit parameters for Sco X-1 and Cyg X-2\label{tab:scocyg}}
\tablehead{ \colhead{} & \colhead{Sco X-1 (NB)} & \colhead{Cyg X-2 (Upper NB)}} 
\startdata
\nh ($10^{21}$ atoms\,cm$^{-2}$)                            	& 1.5$^a$   & 2$^a$   \\
$kT_{\rm disk}$ (keV)                                        		& 1.78$\pm$0.02 &  1.64$\pm$0.02 \\   
$R_{\rm in}$ (km)                                               		& 9.73$\pm$0.18 & 18.4$\pm$0.6  \\  
$kT_{\rm bb}$ (keV)                                          		& 2.64$\pm$0.02 & 2.37$\pm$0.06 \\
$R_{\rm bb}$ (km)                                           		& 2.94$\pm$0.09 &  3.62$\pm$0.15\\
PL index                                                                      & 3.70$\pm$0.05 & 2.4$\pm$0.3\\
$E_{\rm cut}$ (keV)                                         		& --- &  13$\pm$5  \\     
$E_{\rm Gauss}$ (keV)                                                       & 6.60$\pm$0.04 &  6.45$\pm$0.09\\
$\sigma_{\rm Gauss}$ (keV)                                                 & 0.61$\pm$0.08 & 0.91$\pm$0.10\\
norm$_{Gauss}$                                                 & (1.38$\pm$0.17)$\times10^{-1}$ & (1.3$\pm$2)$\times10^{-2}$ \\
$F_{\rm 3-50\,keV}$$^b$  ($10^{-8}$ erg\,\,cm$^{-2}$\,s$^{-1}$)	     &   19.22$\pm$0.03 &   1.57$\pm$0.12     \\      
PL Fraction (\%)    	                                                 & 0.9$\pm$0.4 & 10$\pm$3 \\
$L_{\rm 1-100\, keV}$ ($10^{38}$ erg\,\,s$^{-1}$)       &  2.445$\pm$0.002 &  3.34$\pm$0.17 \\ 
\enddata 
\tablecomments{For $R_{\rm in}$, $R_{\rm bb}$, and $L_{\rm 1-100}$ we assumed distances of 2.8 kpc and 11 kpc for Sco X-1 and Cyg X-2, respectively \citep{brfoge1999,gamuha2008}. For $R_{\rm in}$  we additionally assumed an inclination of 30$^\circ$ for Sco X-1 and  60$^\circ$ for Cyg X-2. For $L_{\rm 1-100\, keV}$ the power-law contribution was only measured between 3 and 100 keV.}
\noindent$^a$ parameter was fixed\\
\noindent$^b$ Absorbed

\end{deluxetable*}

For comparison with \src, we  analyzed \nustar\ data of two other Z sources: Sco X-1 (observation 30001040002) and Cyg X-2 (observation 30001141002,  see also \citet{modepa2018}). The same analysis steps were followed as for \src. In the HIDs of both sources, a clear NB could be identified; for Sco X-1, we extracted a spectrum from the full NB, while for Cyg X-2 we extracted a spectrum from the upper NB (as the lower part was relatively sparsely covered). These spectra had similar signal-to-noise ratios as those of \src, and {we applied the same model and procedure as for \src\ to study the residuals in the 5--8 keV range.} The resulting data/model ratios are shown in Figures \ref{fig:ratios}f and \ref{fig:ratios}g and obvious broad positive residuals around 6.5 keV can be seen for Sco X-1 and Cyg X-2. {  In Table \ref{tab:scocyg} we give the fit parameters to the spectra of Sco X-1 and Cyg X-2, with a Gaussian added to our model to account for the \fe\ line. Note that for Sco X-1, we used a power law without a cutoff. The fit parameters for Sco X-1 and Cyg X-2 are comparable to those found for \src\ on the upper and lower NB, except for the fact that the power-law contribution appears to be higher in \src, especially compared to Sco X-1.}

When fitted with a simple Gaussian, the lines in Sco X-1 and Cyg X-2 have { equivalent widths $\sim$80 eV}. Using a Gaussian with similar { widths} as we find for Sco X-1 and Cyg X-2 ({0.6--}0.9 keV), we obtain 3$\sigma$ upper limits on the equivalent width of less than 10 eV in the 6.4--6.7 keV range for \src. 

We tried to constrain the reflection fraction in our \src\ spectra using  {\tt bbrefl}  \citep[][using solar Fe abundances]{ba2004}, which models the reflection spectrum from a constant-density disk illuminated by a blackbody component.  Since the version of the {\tt bbrefl} model that we used  includes a blackbody component, the blackbody component that  was already part of our model (to account for the boundary layer emission) was removed.  In our fits to the  HB spectrum, in which we found the boundary layer to have the highest temperature, the ionization parameter, $\log(\xi)$, pegged at its highest allowed value of 3.75. The reflection fraction $f_{\rm refl}$ is very low, with a $3\sigma$ upper limit of  $\sim$0.03. The boundary layer temperature was 2.41$\pm$0.02 keV (compared to 2.51$\pm$0.04 keV without the inclusion of the reflection component, see Table \ref{tab:fits}), while the normalization of the {\tt bbrefl} component had a value of 1.98$^{+2.00}_{-0.06}\times10^{-26}$. Spectra from other parts of the Z track showed similar behavior: very low reflection fractions and $\log(\xi)$ values close to or pegged at 3.75. For comparison, using the same model for Sco X-1, we found $\log(\xi)=3.32\pm0.07$ and $f_{\rm refl}=1.0^{+0.8}_{-0.3}$, while for Cyg X-2 we obtained $\log(\xi)=3.66^{+0.01}_{-0.03}$ and  $f_{\rm refl}$ at its maximum allowed value of 5.0.

\section{Discussion}\label{sec:disc}

We have analyzed three \nustar\ observations of the Cyg-like Z source \src. Together, these three observations cover almost the entire Z track of the source, from the HB to the dipping FB. The main result from our analysis is the lack of clear disk reflection features in any of the spectra  along the Z track of \src. This is surprising, given the presence of strong and broad \fe\ lines in \nustar\ spectra of many NS-LMXBs (see Table \ref{tab:overview}), and in particular the Z sources GX 17+2 \citep{lumiba2017}, GX 349+2 \citep{cocami2017}, Cyg X-2 \citep[][this work]{modepa2018}, and  Sco X-1 (this work). A broad Fe line has also been reported in {\it XMM-Newton} spectra of the Z source GX 340+0 \citep{daiadi2009,camiba2010}, a source that is often considered a close  twin of \src. The only two NS-LMXBs without reported reflection features in \nustar\ spectra are GS 1826--238 \citep{chgain2016}, which had a luminosity of $\sim0.13\,L_{\rm Edd}$, and Cen X-4 \citep{chtogr2014}, which was observed in quiescence at 3.8$\times10^{-6}$ $L_{\rm Edd}$.

Based on a study of broad \fe\ lines in 10 NS-LMXBs, \citet{camiba2010} concluded that the boundary layer emission dominates the flux irradiating the inner accretion disk. \src\ shows a hot  and strong boundary layer component in our \nustar\ spectra, reaching temperatures as high as $\sim$2.5 keV on the HB. Our fits also suggest that the inner accretion disk extends close to the neutron star on the HB and NB. Given this combination of parameters one would expect to see \fe\ reflection features on at least part of the Z track of \src. In the following we discuss several possible explanations for the absence of reflection features in the X-ray spectra of \src: 

\begin{enumerate}

{\item As the ionization parameter  $\xi$ increases to a few times $10^4$ one expects the \fe\ line to become weakened and broadened as the result  of Compton scattering \citep{rofayo1999}, making it hard to detect them, even in bright sources. As can be seen from Table \ref{tab:overview}, \src\ is at the top of the luminosity range of NS-LMXBs observed with \nustar, although it should be noted that distances to many of the NS-LMXBs are uncertain.  Given the high $\xi$ values we measured in Sco X-1 and Cyg X-2, a $\xi$ value close to $10^4$ for \src\ might not be unreasonable.  }

 {\item The Fe abundance plays an important role in the overall shape of the reflection component, with a low Fe abundance resulting in a weak or absent \fe\ line \citep{rofayo1999}. However, the lower limit on the Fe abundance reported by \citet{zecode2017}, $>$0.76, rules out a highly subsolar Fe abundance.}
 
 {\item The inclination affects the observed shape of \fe\ lines produced in the inner accretion disk, as the result of relativistic effects, leading to very broad lines in the case of nearly edge-on systems \citep{la1991} that might blend in with the continuum components. However, the high viewing angles required for extreme broadening would likely result in observational effects such as periodic absorption dips or eclipses in the light curves \citep{frkila1987}, neither of which are  seen in \src.}
 
 \item{As the luminosity of NS-LMXBs approaches or exceeds the Eddington luminosity, changes in the accretion geometry are likely to occur, as witnessed by the dramatic changes in the Z tracks of, e.g., XTE J1701--462 and Cir X-1 \citep{frhore2015}. Although the nature of these changes is unknown, it is possible that a combination of changes in the boundary layer and accretion disk result in less efficient illumination of the inner accretion disk at super-Eddington luminosities, causing weaker reflection features. We deem this unlikely, given the presence of  strong \fe\ lines in other super-Eddington Z sources.}
 
\end{enumerate}

Given the high luminosity of \src, we consider high $\xi$ values the most likely explanation for the absence of reflection features in its spectra, although this is not without { potential challenges. First,} why do GX 340+0, a close twin to \src, and Cyg X-2, which had a similar luminosity to \src\ in its \nustar\ observations, both show broad \fe\ lines, while \src\ does not? A proper luminosity comparison of all Z sources would be required to determine how the luminosity of \src\ truly ranks among the Z sources. This, in turn, requires more accurate distances, inclinations, and spectra measured over a wider energy range. {Second, if there are no reflection features from the inner disk because it is too highly ionized, why are they not observed from less ionized regions further out  in the disk? The large thickness of the inner accretion disk at super-Eddington luminosities will likely lead to shielding of the those parts of the disk, thereby limiting the amount of emission from the central source that could produce reflection features. Finally, as discussed in \citet{whlizd1988} and \citet{zdlugi2003} even in the case of a fully ionized disk, the Klein-Nishina process should lead to an observable high-energy break (around a few tens of keV) due to reflection, especially if the illuminating emission extends well beyond that energy. However, since the reflection in high-luminosity NS-LMXBs appears to be dominated by emission originating in the boundary layer \citep{camiba2010}, a high-energy break in the reflected black-body emission from the boundary layer might be difficult to detect. Indeed, we do not detect such a break in the spectra of Sco X-1 and Cyg X-2 that we analyzed either;  both of these spectra show \fe\ lines, indicating that reflection is taking place, and both suggest a high-ionization state of the disk ($\xi\sim3.3-3.7$).}

The absence of reflection features in \src\ may indicate a link with ultra-luminous X-ray sources (ULXs).  We measured 1--100 keV fluxes of (3.0--4.3)$\times10^{38}$($d_{9}$)$^2$ erg\,\,s$^{-1}$, which start to approach the traditionally used lower boundary of ULXs ($10^{39}$ erg\,\,s$^{-1}$). Strong curvature in the 6--8 keV range in {\it XMM-Newton} spectra of some ULXs has been interpreted as a relativistically broadened \fe\ line, but \nustar\ data have shown that reflection models can be ruled out as an explanation of that feature \citep{wahagr2014,rahaba2015}.  As far as we are aware, reflection features have not yet been observed in \nustar\ spectra of ULXs. Perhaps, as we suggest for \src, the inner accretion disks in ULXs are also fully ionized, as the result of their super-Eddington luminosities. In NS-LMXBs the luminosity of \src\ ($\sim$2\,\ledd) possibly represents a critical value for the presence of reflection features (see Table \ref{tab:overview}).

The lack of reflection features in the spectra of \src\ means that we could not perform tests of our continuum model by obtaining independent measurements of the inner disk radius. Although the overall decrease in disk and/or boundary layer temperatures from the HB down to the dipping FB is consistent with earlier spectral studies of \src\ \citep{paebti2005,jachba2009},  disk properties, as well as the \nh\ and the power-law index, remain difficult to constrain without coverage below 3 keV. Simultaneous coverage with {\it NICER} \citep{gearad2016} during future \nustar\ observations of Z sources may provide better constraints on the spectral evolution along their Z tracks.

\acknowledgments

J.H.\ acknowledges financial support  from NASA grant NNX15AV35G. R.L.\ acknowledges funding through a NASA Earth and Space Sciences Fellowship. We thank Sasha Zeegers for discussions on the Fe abundance and \nh\ values of \src{, and Andrzej Zdziarski for his suggestions about the presence of a possible high-energy break}. This research has made use of data and/or software provided by the High Energy Astrophysics Science Archive Research Center, which is a service of the Astrophysics Science Division at NASA/GSFC and the High Energy Astrophysics Division of the Smithsonian Astrophysical Observatory.


\end{document}